\documentclass[conference,a4paper]{APSIPA2025}
\usepackage{amsmath,bm,amssymb}

\usepackage{graphicx}
\usepackage{multirow}
\usepackage{threeparttable}
\usepackage[backend=biber,style=ieee,]{biblatex}
\usepackage{geometry}
\usepackage{fancyhdr}

\fancypagestyle{firststyle}{
  \fancyhf{}
  \fancyhead[C]{2025 Asia Pacific Signal and Information Processing Association Annual Summit and Conference (APSIPA ASC)}
}

\begin{document}

\title{Neural Speech Separation with Parallel Amplitude and Phase Spectrum Estimation}

\author{
\authorblockN{
Fei Liu, 
Yang Ai$^*$,
Zhen-Hua Ling
}

\authorblockA{
National Engineering Research Center of Speech and Language Information Processing,\\ 
University of Science and Technology of China, Hefei, China \\
E-mail: fliu215@mail.ustc.edu.cn,  yangai@ustc.edu.cn, zhling@ustc.edu.cn}
}

\maketitle
\thispagestyle{firststyle}
\pagestyle{fancy}

\begin{abstract}
This paper proposes APSS, a novel neural speech separation model with parallel amplitude and phase spectrum estimation. 
Unlike most existing speech separation methods, the APSS distinguishes itself by explicitly estimating the phase spectrum for more complete and accurate separation. 
Specifically, APSS first extracts the amplitude and phase spectra from the mixed speech signal. 
Subsequently, the extracted amplitude and phase spectra are fused by a feature combiner into joint representations, which are then further processed by a deep processor with time-frequency Transformers to capture temporal and spectral dependencies.
Finally, leveraging parallel amplitude and phase separators, the APSS estimates the respective spectra for each speaker from the resulting features, which are then combined via inverse short-time Fourier transform (iSTFT) to reconstruct the separated speech signals.
Experimental results indicate that APSS surpasses both time-domain separation methods and implicit-phase-estimation-based time-frequency approaches. 
Also, APSS achieves stable and competitive results on multiple datasets, highlighting its strong generalization capability and practical applicability.
\end{abstract}

\section{Introduction}
\renewcommand{\thefootnote}{}
\footnote{$^*$ Corresponding author. This work was funded by the Anhui Province Major Science and Technology Research Project under Grant S2023Z20004, the National Nature Science Foundation of China under Grant 62301521 and the Anhui Provincial Natural Science Foundation under Grant 2308085QF200.}
 \renewcommand{\thefootnote}{\arabic{footnote}}
\addtocounter{footnote}{-1}

In noisy indoor environments such as cocktail parties, multiple people often speak simultaneously, accompanied by background noise. 
Yet, people always seem able to effortlessly separate a target speaker’s voice from the mixture of sound sources and focus solely on that speaker. 
This phenomenon is known as the “cocktail party problem” \cite{cherry1953}.
Speech separation arises from this problem. In real-world scenarios, audio recordings typically contain not only the voice of the primary speaker but also interference from other speakers and background noise. Therefore, the goal of speech separation is to extract the useful speech signal from the corrupted mixture. 
The separated speech can then be used in downstream tasks such as automatic speech recognition \cite{chang2020end}, improving the accuracy and robustness of these systems.
This paper primarily focuses on monaural two-speaker speech separation, which aims to separate the voices of two speakers from a mixture recorded using a single microphone.

In the early stages, researchers use signal-processing-based methods to separate speech \cite{wiener2006new}. 
Under the assumption of known prior distributions for speech and interference, these methods infer the spectral coefficients of speech from the mixed signal to achieve separation. For example, Wiener filtering is an optimal filter that separates speech in the sense of minimizing mean square error \cite{wiener2006new}.
Signal processing methods may perform well under ideal conditions where their assumptions are met; however, in real-world scenarios, these assumptions rarely hold, resulting in a substantial degradation in performance. 
Later, researchers introduce decomposition-based methods, which assume that the sound spectrogram has a low-rank structure and can therefore be represented by a small set of basis vectors \cite{ozerov2011general}. 
However, these methods suffer from poor generalization and degraded performance when the input speech does not match the training conditions. 
For example, the non-negative matrix factorization (NMF) method \cite{virtanen2007monaural} factorizes any non-negative matrix into the product of two non-negative matrices, extracting local basis representations. 
Nonetheless, it struggles to capture deep nonlinear features and involves time-consuming inference, making it impractical for real-time applications.
To overcome these limitations, Wang \MakeLowercase{\textit{et al.}} propose computational auditory scene analysis (CASA) method \cite{wang2006computational}, which simulates the auditory masking mechanism of the human ear by estimating an ideal binary mask to achieve speech separation. 
This approach does not rely on strict prior assumptions and thus offers better generalization. 
However, it heavily depends on the accuracy of pitch estimation, which is challenging in complex environments, limiting its practical applicability.

\begin{figure*}
    \centering
    \includegraphics[width=\linewidth]{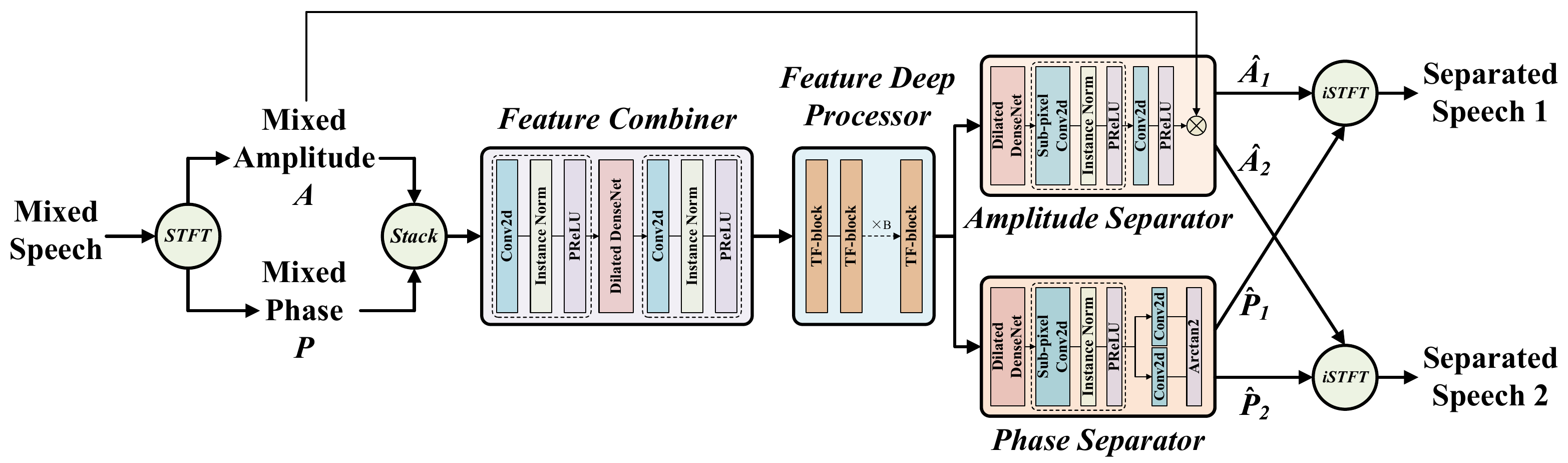}
    \caption{Overview of the proposed APSS. The ``Stack" denotes the spectral concatenation operation and the ``Arctan2" denotes the two-argument arc-tangent function.}
    \label{fig:model}
\end{figure*}

In recent years, with the advancement of deep learning, data-driven neural networks have been widely applied to speech separation. 
Initially, most researchers adopt approaches based on the time-frequency representation of the mixed speech. 
They apply the short-time Fourier transform (STFT) to the mixed speech waveform to obtain its spectra, and then estimate either the amplitude spectrum of each individual source or a corresponding mask \cite{liang2014analysis}. 
For example, early deep clustering methods use ideal binary masks (IBM) for separation \cite{hershey2016deep}.
However, whether estimating the amplitude directly or using a mask, the waveform of each source is ultimately reconstructed by combining the estimated amplitude spectrum with the phase spectrum of the original mixture using the inverse STFT (iSTFT). 
Even with an ideal amplitude spectrum, errors in the phase spectrum impose an upper limit on the accuracy of the reconstructed speech. 
Although phase reconstruction techniques can partially mitigate this issue, accurate phase estimation remains a complex challenge, and current methods still fall short of achieving satisfactory performance.
Some researchers adopt a real-and-imaginary component modeling approach in the time-frequency domain to implicitly model phase \cite{tfpsnet,deepCASA}, achieving promising speech separation results.
Alternatively, some approaches model speech separation directly in the time domain to circumvent the decoupling of amplitude and phase \cite{afrcnn,Convtasnet,dprnn,dptnet,s4m,sandglasset,sepformer,Sudormrf,tasnet,tdanet,Twostep2020,wavesplit}.
For instance, TasNet \cite{tasnet} introduces learnable encoder and decoder modules and estimates a mask from the encoder output to achieve source separation. 
However, the improvement brought by this method is limited, and modeling long sequences in the time domain introduces greater challenges and significantly increases computational cost. Consequently, the lack of direct phase prediction still inevitably impacts speech separation performance.

Therefore, this paper proposes a novel Amplitude-Phase-estimation-based neural Speech Separation model called APSS. 
The APSS adopts a time-frequency domain modeling approach and is primarily composed of a feature combiner, a feature deep processor, an amplitude separator and a phase separator.
First, the model extracts the amplitude and phase spectra of the mixed speech waveform using STFT. 
These spectra are then jointly fed into the feature combiner to produce high-dimensional fusion features. The feature deep processor leverages both time-domain and frequency-domain Transformers to capture temporal and spectral dependencies in the fusion features. 
Next, parallel amplitude and phase separators independently estimate the original amplitude and phase spectrum for each speaker. 
Finally, the iSTFT is applied to reconstruct each separated speech waveform.
By explicitly modeling phase information, APSS effectively mitigates the compensation effect between amplitude and phase \cite{lu2023explicit}.
It also avoids the challenge of modeling long sequences in the time domain. 
The experimental results validate that our proposed APSS outperforms both time-domain speech separation methods and time-frequency domain approaches relying on implicit phase modeling.

This paper is organized as follows.
In Section \ref{sec: propose}, we provide details of the proposed APSS.
In Section \ref{sec: Experiments}, we present our experimental results.
Finally, we give conclusions in Section \ref{sec: Conclusion}.

\section{Proposed Method}
\label{sec: propose}
\subsection{Overview}

\begin{figure*}
    \centering
    \includegraphics[width=0.8\linewidth]{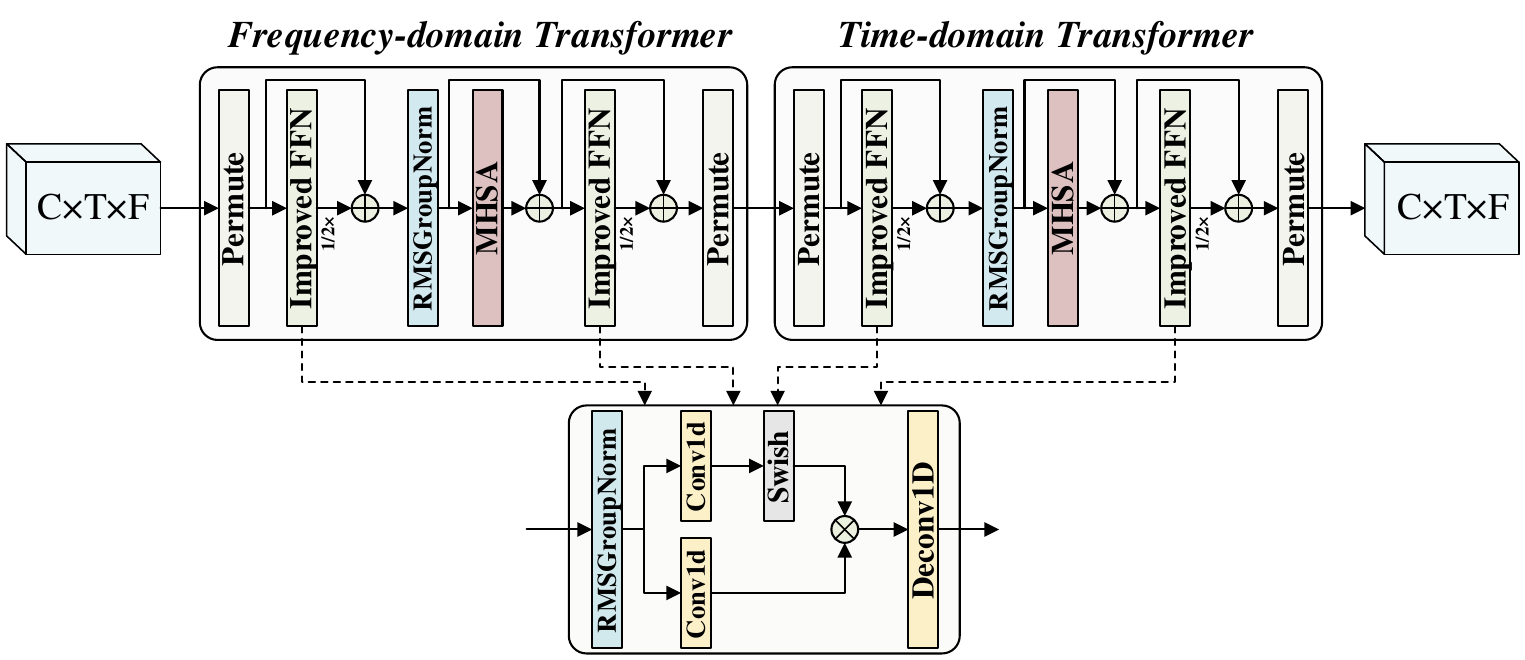}
    \caption{The details of the TF-block employed in APSS.}
    \label{fig:tfblock}
\end{figure*}

The overall architecture of the proposed APSS model is illustrated in Figure \ref{fig:model}.
APSS primarily consists of a feature combiner $\phi_{FC}$, a feature deep processor $\phi_{FDP}$, an amplitude separator $\phi_{AS}$ and a phase separator $\phi_{PS}$, working together to separate monaural speech signals from different speakers in the time-frequency domain.
We denote the mixture of two speech signals ($\bm{x}_{1}, \bm{x}_{2} \in \mathbb R^{L}$) as $\bm{x}=\bm{x}_{1}+\bm{x}_{2} \in \mathbb R^{L}$, where $L$ is the length of the waveform. 
The mixed speech $\bm{x}$ is first transformed using STFT to extract the mixed amplitude spectrum $\bm{A}\in \mathbb R^{T \times F}$ and the mixed phase spectrum $\bm{P}\in \mathbb R^{T \times F}$, where $T$ is the number of frames and $F$ is the number of frequency bins. 
The amplitude and phase spectra are stacked together to form the feature combiner's input $\bm{X}\in \mathbb R^{2\times T\times F}$, which is then processed by the feature combiner to produce a high-dimensional fused feature $\bm{E}\in \mathbb R^{C\times T\times F}$, i.e.,
\begin{gather}
\bm{X}=Stack(\bm{A,P}), \\
\bm{E}=\phi_{FC}(\bm{X}),
\end{gather}
where $C$ denotes the number of feature channels.
The feature combiner effectively exploits the coupling and correlation between amplitude and phase, enhancing the performance of the subsequent separation process.
Then, the fused feature $\bm{E}$ is passed through the feature deep processor to produce deep feature $\bm{S}\in \mathbb R^{C\times T\times F}$, i.e.,
\begin{equation}
\bm{S}=\phi_{FDP}(\bm{E}).
\end{equation}
The feature deep processor effectively captures temporal and spectral dependencies, thereby reducing the complexity of the subsequent separation process.
The deep feature is then fed into two parallel separators: an amplitude separator and a phase separator, which estimate the separated amplitude spectra $\bm{\hat{A}}_1,\bm{\hat{A}}_2\in \mathbb R^{T\times F}$ and the separated phase spectra $\bm{\hat{P}}_1,\bm{\hat{P}}_2\in \mathbb R^{T\times F}$, respectively, i.e.,
\begin{gather}
\bm{\hat{A}}_1,\bm{\hat{A}}_2=\phi_{AS}(\bm{S}), \\
\bm{\hat{P}}_1,\bm{\hat{P}}_2=\phi_{PS}(\bm{S}).
\end{gather}
Although the two separators estimate amplitude and phase independently, their inputs incorporate both amplitude and phase information. This is due to the intrinsic correlation between amplitude and phase, where each can facilitate the prediction of the other.
Finally, the separated speech signals ($\hat{\bm{x}}_{1}, \hat{\bm{x}}_{2} \in \mathbb R^{L}$) are respectively reconstructed via iSTFT, i.e., 
\begin{gather}
\hat{\bm{x}}_{1}=iSTFT(\bm{\hat{A}}_1e^{j\bm{\hat{P}}_1}),\\
\hat{\bm{x}}_{2}=iSTFT(\bm{\hat{A}}_2e^{j\bm{\hat{P}}_2}).
\end{gather}

\subsection{Model Details}
\subsubsection{Feature Combiner}

As shown in Figure \ref{fig:model}, the feature combiner consists of two convolutional blocks and a dilated DenseNet \cite{pandey2020densely}. 
It actually applies dimensional transformation to the input feature $\bm{X}$, producing the fused feature $\bm{E}$ with $C$ channels. 
Specifically, each convolutional block is composed of a 2D convolution layer, an instance normalization layer, and a parametric rectified linear unit (PReLU) activation function. 
The first convolutional block increases the number of channels from 2 to $C$, fully integrating amplitude and phase information. 
The following dilated DenseNet expands the receptive field along the time axis and concatenates the output of each convolutional layer with all preceding layers, enhancing feature reuse and temporal modeling.
Finally, the second convolutional block performs preliminary processing along the frequency axis to produce the final fused feature $\bm{E}$.

\subsubsection{Feature Deep Processor}

The dual-path attention structure has been shown to deliver excellent perfermance in speech separation tasks \cite{dptnet}.
As illustrated in Figure \ref{fig:model}, the feature deep processor consists of $B$ TF-blocks, each of which sequentially captures temporal and spectral dependencies, making it easier to learn the distinctions between different speakers.
As shown in Figure \ref{fig:tfblock}, each TF-block includes a frequency-domain Transformer and a time-domain Transformer, both of which share the same architecture. 
Compared with the standard Transformer, we adopt an improved feed-forward network (FFN), replacing the two linear layers with a one-dimensional convolution layer and a one-dimensional transposed convolution layer. 
Additionally, we substitute the original ReLU activation function with the SwiGLU \cite{shazeer2020glu} activation function to improve the Transformer's performance. 
SwiGLU combines the smoothness of Swish with the gating mechanism of GLU, enabling the model to more effectively learn diverse features from the input data. 
This activation function has been widely adopted in various Transformer architectures \cite{tay2022efficient,touvron2023llama}.
We also introduce an improved normalization method: instead of computing the mean of the input, we apply a root mean square (RMS)-like normalization across grouped features. 
This approach not only simplifies computation but also avoids instability caused by mean estimation, resulting in faster and more stable convergence during training. 
Specifically, each TF-block takes the fused feature $\bm{E}$ with a shape $C\times T\times F$ as input. 
It first applies a dimension transformation to produce an intermediate representation of shape $T\times F\times C$.
This is passed through an FFN and then into a multi-head self-attention layer to capture frequency dependencies, followed by another FFN. After processing through the frequency-domain Transformer, the output is reshaped to $F\times T\times C$ and then passed through the time-domain Transformer to capture temporal dependencies. Finally, the output is reshaped back to $C\times T\times F$, forming the output of the TF-block.

\subsubsection{Amplitude Separator}

As shown in Figure \ref{fig:model}, the amplitude separator consists of a dilated DenseNet, a deconvolutional block, and a mask estimation module. 
The deconvolutional block is structurally similar to the convolutional blocks in the feature combiner, but it replaces the 2D convolution with a 2D sub-pixel convolution \cite{shi2016real}.
The mask estimation module is composed of a 2D convolutional layer followed by a PReLU activation function. 
The deep feature $\bm{S}$ produced by the feature deep processor are passed through the amplitude separator to generate the amplitude masks $\bm{\hat{M}}_1,\bm{\hat{M}}_2\in \mathbb{R}^{T \times F}$ for separated speech signals. 
These masks are then element-wise multiplied with the mixed amplitude spectrum $\bm{A}$ to obtain the separated amplitude spectrum $\bm{\hat{A}_1}$, $\bm{\hat{A}_2}$, i.e.,
\begin{equation}
\bm{\hat{A}}_1=\bm{\hat{M}}_1 \odot \bm{A},
\end{equation}
\begin{equation}
\bm{\hat{A}}_2=\bm{\hat{M}}_2 \odot \bm{A},
\end{equation}
where $\odot$ represents the element-wise multiplication.

\subsubsection{Phase Separator}

As shown in Figure \ref{fig:model}, the phase separator shares a similar structure with the amplitude separator. However, due to the inherent phase wrapping problem, we do not directly reuse the same architecture for both separators. Instead, inspired by \cite{ai2023neural}, we adopt a parallel estimation architecture (PEA) to predict the wrapped phase. 
Specifically, two parallel 2D convolutional branches are used to estimate the pseudo-real and pseudo-imaginary components, which are then activated using the two-argument arctangent function to produce the separated phase spectra $\bm{\hat{P}}_1,\bm{\hat{P}}_2$.

\subsection{Training Criteria}

During training, we use scale-invariant signal-to-noise ratio (SI-SNR) as the loss function for optimization, which is defined as follows:
\begin{equation}
\mathcal L=-20\log_{10}{\frac{\left \| \bm{e}_1 \right \|_{2} \left \| \bm{e}_2 \right \|_{2} }{\left \| \bm{\hat{x}}_{1}-\bm{e}_1 \right \|_{2} \left \| \bm{\hat{x}}_{2}-\bm{e}_2 \right \|_{2}} },
\end{equation}
where
\begin{equation}
\bm{e}_1=\frac{\bm{\hat{x}}_{1}^\top\bm{x}_1}{\left \| \bm{x}_1 \right \|_{2}^{2}}\bm{x}_1, \quad\bm{e}_2=\frac{\bm{\hat{x}}_{2}^\top\bm{x}_2}{\left \| \bm{x}_2 \right \|_{2}^{2}}\bm{x}_2,
\end{equation}
and $\left \| \cdot \right \|_{2}$ denotes L2-norm.
Since the output order of the separation model is inherently ambiguous, we adopt permutation invariant train (PIT) \cite{pit}. 
PIT computes the prediction error for all possible output-target permutations and selects the permutation with the lowest error to guide network training. This process continues until the model converges.

\section{Experiments}
\label{sec: Experiments}
\subsection{Dataset}

We evaluated the performance of the proposed APSS model on two commonly used monaural speech separation datasets, i.e., WSJ0-2Mix \cite{hershey2016deep} and Libri2Mix \cite{librimix}. 
For all datasets, we used the fully overlapping minimum versions and set the sampling rate to 8 kHz.
\begin{itemize}[]
\item {}{\textbf{WSJ0-2Mix}:}
WSJ0-2Mix consists of two-speaker mixed speech, where each mixture is generated by randomly selecting utterances from the corresponding sets and mixing them at a randomly chosen signal-to-noise ratio (SNR) between 0 dB and 5 dB. The dataset contains approximately 30 hours of training data, 10 hours of validation data, and 5 hours of test data.
\item {}{\textbf{Libri2Mix}:}
The Libri2Mix dataset was constructed by randomly selecting speech from two different speakers in the train-100 subset of LibriSpeech \cite{panayotov2015librispeech}, and mixing them with uniformly sampled loudness units relative to full scale (LUFS) to get a mixture at an SNR between -25 and -33 dB. 
The dataset contains approximately 58 hours of training data, 11 hours of validation data, and 11 hours of test data. 
Following previous work \cite{dptnet,tdanet}, we used the clean version of the dataset for our experiments.
\end{itemize}

\subsection{Experimental Setup}

When extracting amplitude and phase spectra from the mixed speech, we set the window length to 16 ms, the hop size to 8 ms, and used 128 FFT points, resulting in 65 frequency bins (i.e. $F=65$). 
In the APSS model, all 1D convolutions and 1D transposed convolutions used a kernel size of 4, while all 2D convolutions used a kernel size of $1\times 3$. And the number of intermediate channels is 128 (i.e. $C=128$). The model employed 6 (i.e. $B=6$) TF-blocks, each with 8 attention heads.
We trained the APSS using the AdamW optimizer on a single Nvidia GeForce RTX 4090 GPU, with $\beta_{1}=0.9,\beta_{2}=0.95$, and a weight decay of 0.01 for 200 epochs.
The initial learning rate was set to 0.001, and a warm-up strategy was applied during the first 4,000 steps. If the validation loss did not improve for two consecutive epochs, the learning rate was reduced by a factor of 0.5.

\subsection{Evaluation Metrics}

To evaluate the performance of the APSS model, we followed common practice in monaural speech separation tasks and adopted widely used objective evaluation metrics, including scale-invariant signal-to-noise ratio improvement (SI-SNRi) and signal-to-distortion ratio improvement (SDRi). 
SI-SNRi measured the improvement in SI-SNR ratio relative to the mixed speech signal. 
It eliminated the influence of volume differences and focused solely on the quality of the target speech reconstruction, providing a fairer assessment of separation performance.
SDRi, on the other hand, measured the improvement in SDR ratio relative to the mixture. 
It accounted for both target speech distortion (e.g., waveform deformation) and residual interference, reflecting the method’s ability to preserve the fidelity of the target speech while suppressing interference.

\subsection{Experimental Results}

\begin{table}[t]
\caption{Experimental results of APSS and baseline speech separation models on WSJ0-2Mix. ``-" denotes unavailable result in original work. ``T" denotes time-domain-based model and ``TF" denotes time-frequency-domain-based model.}
\label{tab:wjs0}
\centering
\begin{tabular}{cccc}
\hline
    Model & Domain & \textbf{SI-SNRi (dB)} & \textbf{SDRi (dB)}  \\ \hline
    TasNet \cite{tasnet} & T & 10.8 & 11.1 \\
    Conv-TasNet \cite{Convtasnet} & T & 15.3 & 15.6 \\
    DeepCASA \cite{deepCASA} & TF & 17.7 & 18.0 \\
    Two-Step CTN \cite{Twostep2020} & T & 16.1 & - \\
    SuDoRM-RF \cite{Sudormrf} & T & 18.9 & - \\
    DPRNN \cite{dprnn} & T & 18.8 & 19.0 \\
    DPTNet \cite{dptnet} & T & 20.2 & 20.3 \\
    WaveSplit \cite{wavesplit}& T & 21.0 & 21.2 \\
    A-FRCNN \cite{afrcnn}& T & 18.3 & 18.6 \\
    SepFormer \cite{sepformer}& T & 20.4 &20.5 \\
    Sandglasset \cite{sandglasset}& T & 20.8 & 21.0 \\
    TFPSNet \cite{tfpsnet}& TF & 21.1 & 21.3 \\
    TDANet \cite{tdanet}& T & 18.5 & 18.7 \\
    S4M \cite{s4m}& T & 20.5 & 20.7 \\ \hline
    APSS & TF & \textbf{21.3} & \textbf{21.5} \\ \hline
\end{tabular}
\end{table}

Table \ref{tab:wjs0} reports the performance of our proposed APSS model compared to several baseline speech separation models on the WSJ0-2Mix dataset.
First, compared with time-domain speech separation models such as TasNet \cite{tasnet} and Conv-TasNet \cite{Convtasnet}, our proposed APSS demonstrated superior performance. 
This provides evidence for the effectiveness of time-frequency-domain approaches. 
Performing deep feature extraction on the amplitude and phase spectra obtained from the mixture may be more beneficial for speech separation than learning directly from the raw waveform.
For two-stage separation methods such as Two-Step CTN \cite{Twostep2020} and DeepCASA \cite{deepCASA}, while they delivered certain performance gains, they introduced greater operational complexity. 
In contrast, APSS not only achieved better performance but also offered a more streamlined and practical solution.
Although TFPSNet \cite{tfpsnet} operates in the time-frequency domain, it lacks explicit phase modeling capability and instead processes the real and imaginary parts of the mixture directly.
As shown in Table \ref{tab:wjs0}, our APSS model achieved a 0.2 dB improvement in both SI-SNRi and SDRi over TFPSNet, suggesting that explicit phase modeling in APSS is indeed effective for speech separation.
Overall, APSS outperformed all other models across the board, further validating that joint amplitude and phase modeling in the time-frequency domain is an effective approach for monaural speech separation.

To validate the generalization ability of our proposed APSS model, we also conducted comparisons with existing separation models on the Libri2Mix dataset. 
We selected the models from Table \ref{tab:wjs0} that were also evaluated on the Libri2Mix dataset reported in their papers as our baselines.
The results are shown in Table \ref{tab:libri}.
As we can see, the proposed APSS consistently outperformed all compared baseline speech separation models. 
This demonstrates that APSS has strong generalization capability and delivers robust performance across different datasets.

\begin{table}[t]
\caption{Experimental results of APSS and baseline speech separation models on Libri2Mix. ``T" denotes time-domain-based model and ``TF" denotes time-frequency-domain-based model.}
\label{tab:libri}
\centering
\begin{tabular}{cccc}
\hline
    Model & Domain & \textbf{SI-SNRi (dB)} & \textbf{SDRi (dB)}  \\ \hline
    Conv-TasNet \cite{Convtasnet} & T & 12.2 & 12.7 \\
    SuDoRM-RF \cite{Sudormrf} & T & 14.0 & 14.4 \\
    DPRNN \cite{dprnn} & T & 16.1 & 16.6 \\
    DPTNet \cite{dptnet} & T & 16.7 & 17.1 \\
    WaveSplit \cite{wavesplit}& T & 16.6 & 17.2 \\
    A-FRCNN \cite{afrcnn}& T & 16.7 & 17.2 \\
    SepFormer \cite{sepformer}& T & 16.5 & 17.0 \\
    S4M \cite{s4m}& T & 16.9 & 17.4 \\ \hline
    APSS & TF & \textbf{17.1} & \textbf{17.6} \\ \hline
\end{tabular}
\end{table}

\subsection{Ablation Studies}

\begin{table}[t]
\caption{Experimental results of ablation studies on WJS0-2mix.}
\label{tab:ablation}
\centering
\begin{tabular}{ccc}
\hline
    Model & \textbf{SI-SNRi (dB)} & \textbf{SDRi (dB)}  \\ \hline
    APSS & \textbf{21.3} & \textbf{21.5} \\
    APSS w/o FC & 17.3 & 17.5 \\
    APSS w/o PEA & 19.9 & 20.0 \\
    APSS w/o AM & 20.0 & 20.1 \\ \hline
\end{tabular}
\end{table}

Next, we conducted three ablation experiments to evaluate the contributions of different modules in APSS, as shown in Table \ref{tab:ablation}. 
First, we replaced the feature combiner with a simple 2D convolution for feature extraction (i.e., APSS w/o FC). 
The results show a noticeable drop in SI-SNRi and SDRi, indicating that effective feature combination helps extract richer and deeper information, which benefits the subsequent separation. 
Inspired by \cite{ai2023neural}, we used the PEA which included two parallel 2D convolutions combined with a two-argument arctangent activation in the phase separator of APSS to address the phase wrapping issue. 
In the ablation study, we replaced the PEA with a single 2D convolution (i.e., APSS w/o PEA). 
The results show that both SI-SNRi and SDRi dropped by more than 1 dB.
This suggests that imprecise phase estimation has a substantial negative impact on the separation performance of APSS.
In addition, we ablated the use of an amplitude mask to examine its impact on separation performance (i.e., APSS w/o AM). 
The results show that directly predicting the amplitude spectrum harms performance by reducing estimation accuracy, which agrees with findings in some speech enhancement studies \cite{lu2023explicit}.

\section{Conclusion}
\label{sec: Conclusion}
In this paper, we propose a novel time-frequency-domain neural speech separation model based on amplitude and phase estimation, named APSS. 
The APSS explicitly models both amplitude and phase through a feature combiner, a feature deep processor, and parallel amplitude-phase separators, effectively leveraging the coupling and correlation between amplitude and phase. 
This approach introduces a new paradigm for speech separation tasks. 
Experimental results demonstrate that APSS outperforms existing time-domain and implicit-phase-estimation-based time-frequency-domain speech separation baseline models. 
In future work, we plan to further enhance separation performance of APSS and extend the amplitude-phase estimation framework to multi-speaker speech separation tasks.

\printbibliography

\end{document}